\pdfoutput=1

\documentclass[aps,prd,amsmath,superscriptaddress,%
               nobibnotes,nofootinbib,preprintnumbers,
               onecolumn,12pt,tightenlines]{revtex4}

\usepackage{graphicx}

\usepackage{ifthen}

\graphicspath{{figures/}}

\usepackage[normalem]{ulem}
\usepackage[usenames]{color}
\definecolor{DarkRed}{rgb}{0.5,0.0,0.0}
\definecolor{DarkGreen}{rgb}{0.0,0.5,0.0}
\definecolor{DarkBlue}{rgb}{0.0,0.0,0.5}
\definecolor{Brown}{cmyk}{0.0,0.8,1,0.6}
\definecolor{Magenta}{rgb}{1.0,0.0,1.0}
\definecolor{DarkMagenta}{rgb}{0.5,0.0,0.5}

\newcommand{\eg}{\textit{e.g.}}

\newcommand{\ifmulticol}[2]{%
  \ifthenelse{\lengthtest{1.9\columnwidth<\textwidth}}{#1}{#2}%
}

\newcommand{\gae}{%
  \ensuremath{\lower 2pt \hbox{%
    $\, \buildrel {\scriptstyle >}\over {\scriptstyle \sim}\,$}%
    }%
  }
\newcommand{\lae}{%
  \ensuremath{\lower 2pt \hbox{%
    $\, \buildrel {\scriptstyle <}\over {\scriptstyle \sim}\,$}%
    }%
  }

\newcommand{\Enr}{\ensuremath{E_{nr}}}

\newcommand{\dRdE}{\ensuremath{\frac{dR}{dE}}}

\newcommand{\dRdEnr}{\ensuremath{\frac{dR}{dE_{nr\!\!\!\!\!}}\;}}

\newcommand{\mchi}{\ensuremath{m_{\chi}}}
\newcommand{\rhochi}{\ensuremath{\rho_{\chi}}}
\newcommand{\nchi}{\ensuremath{n_{\chi}}}

\newcommand{\vmin}{\ensuremath{v_\textrm{min}}}
\newcommand{\vmp}{\ensuremath{\overline{v}_0}}
\newcommand{\vrot}{\ensuremath{v_\mathrm{rot}}}

\newcommand{\vesc}{\ensuremath{v_\textrm{esc}}}

\newcommand{\bv}{\ensuremath{\mathbf{v}}}  

\newcommand{\sigmaSI}{\ensuremath{\sigma_{\mathrm{SI}}}}

\newcommand{\sigmapSI}{\ensuremath{\sigma_{\mathrm{p,SI}}}}

\newcommand{\mup}{\ensuremath{\mu_{\mathrm{p}}}}

\usepackage[hyperfootnotes=false]{hyperref}

\begin{document}


\preprint{MCTP--12-14.}


\title{New Dark Matter Detectors using DNA or RNA for Nanometer Tracking }

\author{Andrzej Drukier}
\email[]{adrukier@gmail.com}
\affiliation{
 BioTraces Inc., 5660 Oak Tanager Ct., Burke, Va. 22015}

\author{Katherine Freese}
\email[]{ktfreese@umich.edu}
\affiliation{
 Michigan Center for Theoretical Physics,
 Department of Physics,
 University of Michigan,
 Ann Arbor, MI 48109}
\affiliation{
Physics Department,
Caltech,
Pasadena, CA 91101}

\author{Alejandro Lopez}
\email[]{aolopez@umich.edu}
\affiliation{
 Michigan Center for Theoretical Physics,
 Department of Physics,
 University of Michigan,
 Ann Arbor, MI 48109}

\author{David Spergel}
\email[]{dns@astro.princeton.edu}
\affiliation{
Department of Astronomy
Princeton University
Princeton, NY 08544}

\author{Charles Cantor}
\email[]{ccantor@sequenom.com}
\affiliation{SEQUENOM, Inc.,
3595 John Hopkins Court,
San Diego, CA 92121}

\author{George Church}
\email[]{gchurch@genetics.med.harvard.edu}
\affiliation{
Department of Genetics,
Harvard University,
Boston, MA 02115}
 
\author{Takeshi Sano}
\email[]{sanotakeshi@gmail.com}
\affiliation{
 DiThera, Inc.,
 San Diego, CA 92121}

\date{\today}



\begin{abstract} 


Weakly Interacting Massive Particles (WIMPs) may constitute most of the matter in the Universe.  
The ability to detect the directionality of recoil nuclei will considerably facilitate detection of WIMPs by means of ``annual modulation effect" and ``diurnal modulation effect".
 Directional sensitivity requires either extremely large gas (TPC) detectors or detectors with a few  nanometer spatial resolution. 
 
In this paper we propose  a novel type of dark matter detector:
 detectors made of DNA or RNA could provide nanometer resolution for tracking, an energy threshold of 0.5 keV, and can operate at room temperature.
When a WIMP from the Galactic Halo elastically scatters off of a nucleus in the detector, the recoiling nucleus then
 traverses hundreds of strings of single stranded nucleic acids (ssNA) with known base sequences and severs ssNA strands along its trajectory.
The location of the break can be identified by amplifying and identifying the segments of cut ssNA using techniques well
known to biologists.  Thus the path of the recoiling nucleus can be tracked to nanometer accuracy.  
 In one such detector concept, the
 transducers are nanometer-thick Au-foils of 1m$\times$1m, and the direction of recoiling nuclei is measured by ``NA Tracking Chamber" 
 consisting of ordered array of ssNA strands. Polymerase Chain Reaction (PCR) and ssNA sequencing  are used to read-out the detector. 
 The proposed detector is smaller and cheaper than other alternatives: 1 kg of gold and $0.1$ to 4 kg of ssNA (depending on length and strand density), packed into 0.01m$^3$, can be used
 to study 10 GeV WIMPs. A variety of other detector target elements could be used in this detector to optimize for different WIMP masses and to identify
 WIMP properties. 
 By leveraging advances in molecular biology, we aim to achieve about 1,000-fold better spatial resolution  than in conventional WIMP detectors at reasonable cost. 

\end{abstract} 

\maketitle


\section{\label{sec:Intro} Introduction}

The Milky Way, along with other galaxies, is well known to be
encompassed in a massive dark matter halo of unknown composition.  
Only 5\% of the Universe consists of ordinary atomic matter, while the
remainder is 23\% dark matter and 72\% dark energy \cite{Bennett:2013,Planck:2014}.  Identifying the
nature of this dark matter is the longest outstanding problem
in modern physics.  Leading candidates for this dark matter are
Weakly Interacting Massive Particles (WIMPs), a generic class of
particles that includes the lightest supersymmetric particle.
These particles undergo weak
interactions and their expected masses range from 1~GeV to 10~TeV.
These particles, if present in thermal equilibrium in the early
universe, annihilate with one another so that a predictable number of
them remain today.  For a wide range of parameters, the relic density of these particles is found to be
roughly in agreement with
the value measured by the Wilkinson Microwave Anisotropy Probe (WMAP) and Planck satellite.

Thirty years ago, Refs. \cite{Drukier:1983gj, Goodman:1984dc} first proposed
that the most efficient laboratory mechanism for detecting weakly interacting particles, 
including WIMPs, is via coherent scattering with nuclei.  Soon after \cite{DFS} computed
detection rates in the context of a Galactic Halo of WIMPs. 
Then development of ultra-pure Ge detectors permitted the first limits on WIMPs  
\cite{Ahlen:1987mn}.  Since that time, a
multitude of experimental efforts to detect WIMPs has been underway, with some of them currently claiming detection. 
The basic goal of direct detection experiments is to measure the energy deposited when weakly interacting particles 
scatter off of nuclei in the detector, depositing 1-10 keV in the nucleus.  Numerous collaborations worldwide 
have been searching for WIMPs using a variety of techniques to detect the nuclear recoil.
The most difficult aspect of these experiments is background rejection. To avoid cosmic rays (CR), the experiments are placed deep underground.
Yet radioactive backgrounds persist; fast neutrons produced by CR are particularly difficult to differentiate from WIMPs. 
Important tools in isolating a WIMP signal are the annual and diurnal modulations (AME and DME)
that would be expected for WIMPs but not for backgrounds.

\medskip\noindent
{\bf Annual Modulation Effect:}

Three of us showed in 1986 that the count rate in WIMP direct detection experiments will experience an
annual modulation \cite{DFS,Freese:1987wu} as a result of
the motion of the Earth around the Sun: the relative velocity of the
detector with respect to the WIMPs depends on the time of year.  Thus
the count rate in detectors should change with a cosine dependence on
time.  
During the past ten years the DAMA experiment \cite{DAMA} has observed such
 an annual modulation.  This experiment consists of a large number of NaI
crystals situated in the Gran Sasso Tunnel and currently reports a 9$\sigma$ detection.
The CoGeNT experiment \cite{COGENT}, made of germanium, also reported
annual modulation.  A third experiment, dilution-refrigerator based CRESST-II \cite{CRESST}, 
also announced count rate above expected background. The CDMS-Si experiment claimed three events consistent with low-mass WIMPs \cite{CDMS-Si}.
There has been much discussion as to whether or not these experiments may be consistent with the
same WIMP parameter range, {\it e.g.} \cite{Kelso:2011gd, Fox:2011px}
Yet CDMS II sees no annual modulation \cite{Ahmed:2012vq}, and CDMS II \cite{Ahmed:2010wy} and SuperCDMS \cite{SuperCDMS} find null results. As well, the measurements from the Xenon based dark matter detectors, XENON \cite{Angle:2011th} and LUX \cite{LUX}, are in tension with the positive results of other experiments. The situation
is perplexing.

\medskip\noindent
{\bf Diurnal Modulation Effect:}  

A major step forward in the field of direct detection would be the development of 
detectors with directional capability ~\cite{spergel}, i.e., the capability to determine which direction
the WIMP came from.  As a result of the elastic scattering of WIMP off of a nucleus
in the detector, the nucleus gets kicked in a forward direction.
Thus by determining the track of the nucleus one could identify the direction of the incoming WIMP (Figure 1).
The WIMP flux in the lab
frame is peaked in the direction of motion of the Sun (which happens to be towards
the constellation Cygnus).  Hence the recoil spectrum for most
energies should be peaked in
the direction opposite to this.  The event rate in the
backward direction is expected to be $\sim 10$ times larger than that in the forward
direction~\cite{spergel,gondolo}.  A directional detector which could measure the direction of the recoiling nuclei
from the interaction is required to detect this 'head-tail' asymmetry.   Given the capability of ascertaining this asymmetry, 
the statistical requirements to show a WIMP
detection would only require $\sim$ 30-100 WIMPs \cite{copi:krauss, ck2, pap1}.  
In a second generation of directional detection
experiments, the measurement of the diurnal variation of the count rate due to the daily rotation of the Earth could provide further information. 
Measurements of both the annual and diurnal modulations could then provide a "smoking gun" for the existence of  WIMPs. 
In addition, any galactic substructure in the WIMP density, such as tidal streams, could show up as spikes
coming from one particular direction in a directional detector.

\begin{figure}[diurnal]
\includegraphics[width=0.48\textwidth,height=0.34\textheight]{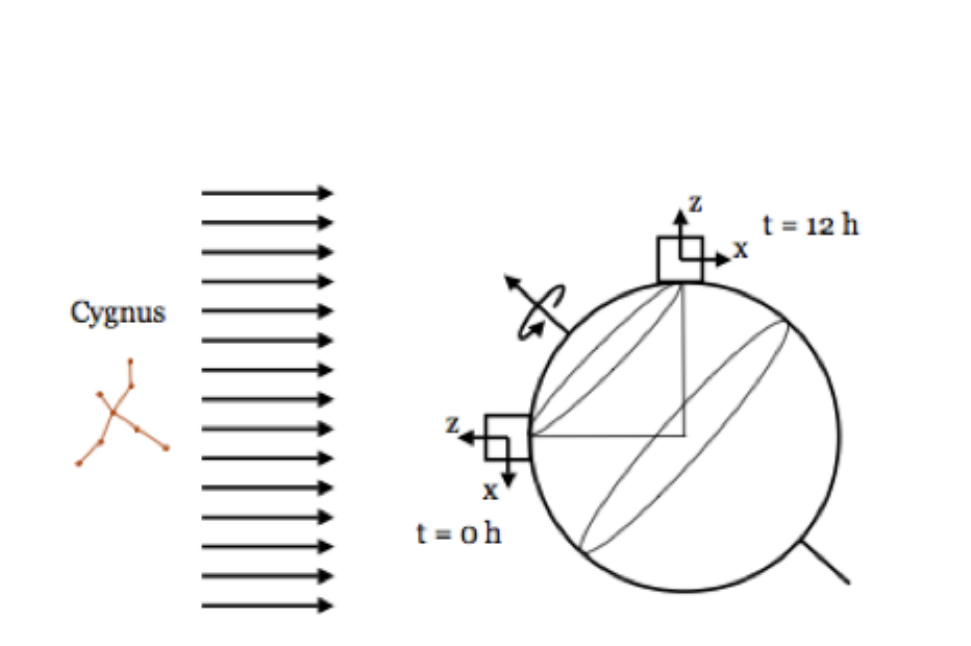}
\caption{Diurnal modulation of WIMPs:  the Sun orbits around the Galactic Center (in a direction that happens to be towards the constellation Cygnus),
 therefore experiencing a
WIMP wind, for which the orientation relative to the laboratory frame depends on the rotation of the earth, and hence time of day.}
\label{fig6}
\end{figure}

\medskip\noindent
{\bf Limitations of existing detectors:}

 The goal is to obtain the track of the recoiling nucleus after it has been hit by a WIMP.  Yet in existing detectors the track length is much
 shorter than the resolution of the detectors. The length of the track of the recoiling nucleus is predicted by Lindhard Theory \cite{Lindhard}. The range of recoiling nuclei is super-short, often below 10 nm, while
existing detectors have spatial resolution of a few microns. 
In both typical solid state detectors as well as liquid xenon detectors, the range is 100 times shorter than the spatial resolution.
As a consequence, in prior designs of "directional detectors", the density of the detectors must be brought low enough  to increase the recoil range. 
For example, it is proposed to use Xe gas pumped to 0.1 Atmosphere \cite{Alner:2005, ahlen, Battat:2010ip}.
A difficulty of this proposal is that such a huge volume of gas must be placed underground and shielded against radioactivity. 

\medskip\noindent
{\bf Polynucleotide based detector:}

 In this paper we describe a smaller and less expensive alternative: detectors made of DNA or RNA may provide nanometer resolution for tracking, 
 energy threshold below 0.5 keV, and can operate at room temperature. 
 One implementation consists of a large number of thin foils of gold (Au) with strings of ssNA hanging down from them as shown in Figure 2. In this paper we take gold to be the target material of the detector; but a variety of other materials could also be used instead, as detailed in the Appendix.
   For simplicity, however, we will call it Au/ssNA detector even if any metal can be used.
 The NA strands all consist of identical sequences of bases (combinations of A,C,G,T), with an order that is well known.
  An incoming WIMP from the Halo of our Galaxy strikes one of the gold nuclei and knocks it out of the film with $\sim$ 10 keV
of energy.  The Au nucleus traverses a few hundred NA strands before stopping.  Whenever it hits the NA, it has a high probability of
severing the ssNA strand.
  The cutoff segment of NA falls down onto a capture foil and is periodically removed.
 The locations of the breaks are easy to identify via a plurality of NA sequencing techniques:  the broken segments can
be copied using Polymerase Chain Reaction (PCR), thus amplifying the signal a billion fold. It can be sequenced with single base accuracy, i.e. $\sim$ nm precision.
Thus the path of the recoiling nucleus can be tracked to nanometer accuracy.  
More details of this particular detector design are presented below.  Alternative detector designs may be implemented instead, but the important new development 
is the idea of using NA in lieu of more conventional detector materials to provide thousand-fold better tracking resolution, so that directionality of the WIMPs can be determined.

There are many advantages to this new technology of using NA:
\begin{enumerate}
  \item Nanometer spatial resolution enables directional detection; 
  \item Operates at room temperature;
  \item Low energy threshold of less than 0.5 keV, allowing for study of low mass $<10$GeV WIMPs;
  \item Flexibility of materials:  One may choose from a variety of elements with high atomic mass ({\it e.g.} Au) to maximize the spin-independent scattering rate.  Given a variety
  of materials one can also extract information about the mass and cross section of the WIMPs;
  \item One can also select materials with high spin to maximize spin-dependent interaction rate;
\item Signal may be amplified by a factor of $10^9$ by using PCR;
\item Excellent background rejection, by using dE/dx in vertex and $> 10^{16}$ physical granularity of the detector,
i.e. there are $10^{16}$ voxels in a (1m)$^3$ detector.
    \end{enumerate}

The nanometer tracking described in this paper may have many uses beyond dark matter detection as will be studied in future work.

\section{\label{sec:DMDetection} Dark Matter Detection}

WIMP direct detection experiments seek to measure the energy deposited
when a WIMP interacts with a nucleus in a detector.
If a WIMP of mass $\mchi$ scatters elastically from a nucleus of mass
$M$, it will deposit a recoil energy $\Enr = (\mu^2 v^2/M)(1-\cos\theta)$,
where $\mu \equiv \mchi M/ (\mchi + M)$ is the reduced mass of the
WIMP-nucleus system, $v$ is the speed of the WIMP relative to the
nucleus, and $\theta$ is the scattering angle in the center of mass frame. Note that the maximum energy recoil is given when the scattering angle in the center of mass is $\theta= \pi$:
\begin{equation}\label{Emax}
E_{max}=2\frac{\mu^2 v^2}{M}.
\end{equation}
The differential recoil rate per unit detector mass, typically given in
units of cpd\,kg$^{-1}$\,keV$^{-1}$ (where cpd is counts per day), can
be written as:
\begin{equation}\label{eqn:dRdEnr}
  \dRdEnr
    = \frac{\nchi}{M} \langle v \frac{d\sigma}{dE} \; \rangle
       = \frac{1}{2 M \mu^2} \, \sigma(q) \, \rhochi \eta(\vmin(\Enr),t) \, ,
\end{equation}
where  $\nchi = \frac{\rhochi}{\mchi}$ is the number density of WIMPs, with $\rhochi$ the local dark matter mass density;
 $q = \sqrt{2 M \Enr}$ is the momentum exchange in the
scatter; $\sigma(q)$ is an effective scattering cross-section;
$  \eta(\vmin,t) = \int_{v > \vmin} d^3v \, \frac{f(\bv,t)}{v}$
is the mean inverse velocity with $f(\bv,t)$  the (time-dependent) WIMP velocity distribution; and
$  \vmin = \sqrt{\frac{M \Enr}{2\mu^2}}$ is
the minimum WIMP velocity that can result in a recoil energy $\Enr$.
More detailed reviews of the dark matter scattering process and direct
detection can be found in Refs.~\cite{Primack:1988zm,Smith:1988kw,
Lewin:1995rx,Jungman:1995df,Bertone:2004pz}.

The typical energy transferred to the nucleus in a scattering event is from 1 to 50 keV. 
Typical count rates in detectors are less than 1 count per kg of detector per day. 
Over the past twenty five years a variety of designs have been developed to detect WIMPs. 
They include detectors that measure scintillation; ionization; and dilution-refrigerator based calorimeters in which the total energy deposed is measured by means of 
a phonon spectrum.  Current detector masses range in size up to 100 kg (e.g. XENON-100 and LUX).
The plan for the next generation of detectors is to reach one tonne.

A major concern in all WIMP detectors is backgrounds. To eliminate spurious events from CR, the detectors must be placed deep underground 
( $>$ 2,000 m of water equivalent). Yet radioactive backgrounds remain and must be eliminated. Thus the experimental determination of 
annual and/or diurnal modulation is a crucial test of the WIMP origin of any events observed in the detector, as most backgrounds should not
exhibit the same time dependence.

\medskip\noindent
{\bf Particle Physics:  WIMP/nucleus cross sections:}

For a supersymmetric (SUSY) neutralino and many other WIMP candidates, the dominant
WIMP-quark couplings in direct detection experiments are the scalar and
axial-vector couplings, which give rise to spin-independent (SI)
and spin-dependent (SD) cross-sections for elastic scattering of a
WIMP with a nucleus, respectively.
 SI scattering is typically taken to be
\begin{equation} \label{eqn:sigmaSI2}
  \sigmaSI = \frac{\mu^2}{\mup^2} A^2 \, \sigmapSI \, ,
\end{equation}
where $A$ is the atomic mass of the nucleus, $\mup$ is the WIMP-proton reduced mass and $\sigmapSI $ is the SI scattering cross section of WIMPs with protons.
For large momentum transfer, this relation is multiplied by a form factor correction to account for the
sensitivity to the spatial structure of the nucleus.
Since the SI cross-section grows rapidly with nuclear mass, direct detection experiments often use heavy
nuclei to increase their sensitivity to WIMP scattering.

Spin-dependent (SD) WIMP-nucleus interactions depend on the spin of the nucleus.  Most nuclei have
equal numbers of neutrons and protons so that there is no SD contribution;  specific nuclei
must be chosen in experiments to search for nonzero SD couplings.  
SD scattering is often of lesser significance than SI scattering in direct detection experiments
 for the heavy elements used in most detectors due to the extra $A^2$ coherence factor
 in the cross section.

\medskip\noindent
{\bf Astrophysics: Velocity Structure of the Galactic Halo:}

The velocity distribution $f(\bv)$ of dark matter particles in the
Galactic Halo is crucial to their signals in dark matter detectors (as first stressed by \cite{DFS}).
The dark matter halo in the local neighbourhood is likely to be
composed mainly of a smooth, well mixed (virialised) component with an
average density  $\rhochi \approx 0.4$~GeV/cm$^3$. The simplest model
of this smooth component is the Standard Halo Model (SHM),
 a spherically symmetric nonrotating isothermal sphere with an isotropic, Maxwellian velocity
distribution characterized by an rms velocity dispersion $\sigma_v \sim 290$ km/sec; the distribution 
 is truncated at  escape velocity $\vesc \sim$ 600 km/sec.   The resultant count rates in direct
 detection experiments due to the SHM were first discussed in \cite{DFS}. 

In addition to the smooth component of the Galaxy, the formation of the
Milky Way via merger events throughout its history leads to significant
structure in both the spatial and velocity distribution of the dark
matter halo. 
The dark matter affiliated with any of these substructures 
(tidal streams of material, subhalos, clumps, caustics, or debris flow) 
located in the  Solar
neighborhood will  affect count rates as well as the phase and amplitude of annual modulation in experiments\footnote{
For example the Sagittarius stream
l~\cite{Freese:2003na,Freese:2003tt} could give an increase in
the count rate in detectors up to a cutoff energy, leading to an annually modulated steplike
feature in the energy recoil
spectrum.   The stream
should also stand out clearly in directional detection experiments
(capable of determining directionality of the incoming WIMPs).}.

\medskip\noindent
{\bf Annual Modulation:}

The smooth component of the halo is essentially non-rotating, while the
Sun moves with the disk and rotates about the center of the galaxy
at a speed $\vrot \sim$ 245 km/sec ~\cite{Bovy2009}.
The halo thus exhibits a bulk motion relative to Earth. One can think of
this phenomenon as the Earth moving into a ``wind" of WIMPs.  
The relative velocity between the WIMPs and the detector plays an important role in detection rates.

This relative velocity experiences two types of modulation:  annual and diurnal.  These
can be very important in proving that any detected signal is in fact due to WIMPs rather than background.
Three of us predicted that, due to the motion of the Earth in orbit about the Sun, the dark matter
velocity distribution as seen by a detector on Earth should undergo a yearly
variation, leading to an annual variation in the recoil rate in the
detector \cite{DFS, Freese:1987wu}.   In many cases, the
 annually modulating recoil rate can be approximated by
\begin{equation} \label{eqn:dRdES}
  \dRdE(E,t) \approx S_0(E) + S_m(E) \cos{\omega(t-t_0)} ,
\end{equation}
with $|S_m| \ll S_0$,
where $S_0$ is the time-averaged rate, $S_m$ is referred to as the
modulation amplitude (which may, in fact, be negative),
$\omega = 2\pi$/year and $t_0$ is
the phase of the modulation.
For the Standard Halo Model, 
the maximum count rate is on June 1 and the minimum on December 1. 
  The modulation is only a few percent of the average count rate.
        Thus, a large number of events are required to observe a
        modulation of the rate in a detector.
We note that, for low enough energy recoils, the typical WIMP is moving in the opposite direction
and the phase of the modulation reverses (the signal is maximized in December); once the crossover energy of this phase reversal is measured
it can be used to determine the WIMP mass\footnote{$x_p = 0.89$ is the value
of $x$ at which the phase of the modulation reverses, where
$x \equiv \vmin/\vmp$ with $\vmin \propto \sqrt{E}$). }.

The reason that annual and diurnal modulation are so powerful as a ``smoking gun'' for
dark matter is that most background signals, \eg\ from radioactivity in
the surroundings, are expected to be isotropic and not modulating with the same
time dependence as the WIMPs.

\medskip\noindent
{\bf Current Experimental Status for Direct Detection:}

In the past decade, a host of direct detection experiments using a variety of different detector materials and designs
 have reported unexplained nuclear recoil signals which could be due to WIMPs. 
Detection of annual modulation has now been claimed by the DAMA and, more
recently, CoGenT experiments.
The Italian Dark Matter Experiment, or DAMA \cite{DAMA}, consists of 250 kg of radio pure NaI scintillator situated in the Gran Sasso Tunnel underneath
the Apennine Mountains near Rome, and became
the first direct detection experiment to observe a
positive signal.  
The group now has accumulated 1 ton-yr of data over the past decade and
  finds an 8.9 $\sigma$ annual modulation with the correct phase and spectrum to be consistent with a dark matter signal.
 Recently CoGeNT \cite{COGENT}, consisting of Germanium, also
claim to see annual modulation of the signal with the correct phase to be consistent with WIMPs, and together with a third
CRESST-II \cite{CRESST} experiment, could be seeing $\sim$10 GeV WIMPs. Positive candidate signals were also seeing in CDMS-Si \cite{CDMS-Si}, which agrees with a $\sim$10 GeV WIMP mass. Yet, other experiments, notably CDMS-Ge \cite{Ahmed:2012vq, Ahmed:2010wy} and SuperCDMS \cite{SuperCDMS} find null results. The results from Xenon based detectors, XENON \cite{Angle:2011th,Angle:2013} and LUX \cite{LUX}, also appear to conflict with these positive signals.   
Many direct detection experiments are either currently running or gearing up to do so,
 and we can expect more data soon. 
 
In the past few years the cross-sections that have been reached by detectors have improved by two orders of magnitude; over the next few years
another two orders of magnitude should be reached.  The next generation of detectors after the current ones will be one tonne in mass or directional.  
A review of the theory and experimental status of annual modulation can be found in \cite{Freesereview}.

\medskip\noindent
{\bf Diurnal Modulation:}

Our motion with respect to the Galactic rest frame also produces a
diurnal modulation of the event rate.  Due
to the motion of the Sun around the Galactic Center we are moving into a "wind" of WIMPs.
As shown in Figure 1, the daily rotation of the Eath then introduces a modulation in recoil angle as measured
in the laboratory frame \cite{spergel}.  The WIMP count rate then is expected to modulate with the time of day.
Measurement of the diurnal modulation would require directional detection capabilities discussed in this paper.  
An ideal detector could reject isotropy of WIMP recoils
with only of order 30-100 events~\cite{copi:krauss,ck2,pap1}.  Most, but not
all, backgrounds would produce an isotropic Galactic recoil
distribution.  
An anisotropic Galactic recoil
distribution would therefore provide strong, but not conclusive,
evidence for a Galactic origin of the recoils. 
Roughly 30 events would
be required for an ideal (no background) detector to confirm that the peak recoil
direction coincides with the inverse of the direction of Solar motion, hence
confirming the Galactic origin of the recoil events~\cite{billard,pap5}.  
Realistically about 100 events consistent with diurnal modulation will be required
for WIMP detection (far less than without this modulation).

Measurement of the diurnal modulation would determine the
direction of the WIMP wind, which could then be compared with an annual modulation signal
found in a different experiment.   If the annual modulation of the signal is dominated by the smooth halo, 
then the wind direction (obtained from the directional experiments) should predict the time of year when
the event rate in direct detection experiments peaks; i.e. the time of year when the Earth 
moves the most quickly into the wind would be the time of peak signal.
 If, on the other hand, the phase of the modulation does not match up with the
wind direction found by the directional detectors, then
one would suspect that the WIMP interpretation of the experiments might be wrong\footnote{Alternatively 
 this discrepancy might be an indication of additional
halo components such as streams, which 
could change the phase of the annual modulation.  In principle comparison between the 
wind direction and the phase of the modulation could teach us about the structure of the dark matter
in our halo.}.

\section{\label{sec:Future}{Directional Detectors}}

Ref.~\cite{ahlen}
reviews the status of one type of prototype directional detection experiments.  Current
designs require the detector material to be gaseous
 in order to produce long enough tracks compared to
the spatial resolution of the detector and will thus require volumes $\sim 1000\  {\rm m}^3.$

In this paper we instead propose the use of DNA or RNA as a detector material that can provide nanometer resolution tracking.  
Figure 2 illustrates an example of a novel detector design, consisting of a thin (5-10 nanometer thick) film of metal (e.g. gold or tungsten) with
strings of single stranded nucleic acids (ssNA) hanging down from it. In Figure 2 we show gold as our detector's target material. It should be noted that a variety of other elements could also be used as target materials. In particular, we could have a film of target material like tungsten on top of a monolayer thin sheet of gold (see the Appendix for details on other possible target elements). For simplicity, in this paper we will use gold as an example for the proposed detector's target element. The monolayer of gold is used as a ceiling for the ordered ssNA to hang down from. The NA strands are all identical in length, with an order of bases that is well known.
However, they may be terminated with �forensic fragment� of say 100-mer.

The basic idea is the following: An incoming WIMP from the Halo of our Galaxy strikes one of the gold nuclei and knocks it out of the film with $\sim$ 10 keV
of energy.  The Au nucleus moves forward into the strands of NA, traverses hundreds of these strands, and whenever it hits one, breaks the ssNA.  A segment
of NA falls down onto a ``capture foil" below.  Periodically ({\it e.g}. once an hour) the fallen segments are scooped up.  The locations of the breaks can be identified:  the strands can
be copied using Polymerase Chain Reaction (PCR), thus amplifying the signal a billion fold; then NA sequencing provides the location of the broken NA.
Since the NA base units are at most $\sim$0.7 nm apart (when fully stretched),
the resulting detector resolution in the z-direction is nanometer.  Thus the track of the recoiling nucleus may be obtained
with nanometer accuracy.

\begin{figure}[gold]
\includegraphics[width=0.5\textwidth,height=0.4\textheight]{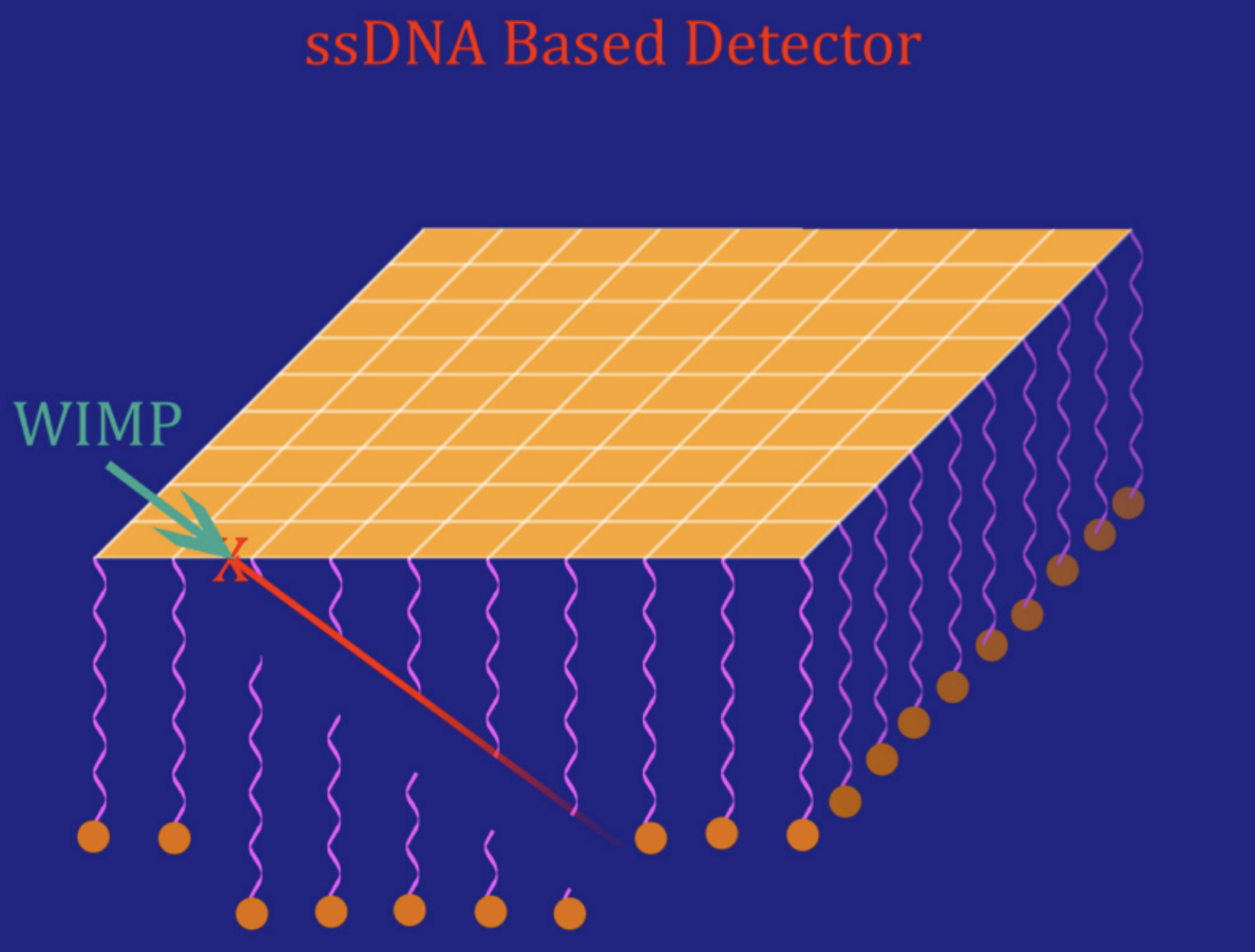}
\caption{ssNA/Au Tracking Chamber:  A WIMP from the Galaxy scatters elastically with a gold nucleus situated in a thin gold foil. The recoiling Au nucleus traverses hanging
strings of single stranded NA, and severs any ssNA it hits.  The location of the breaks can be found by amplifying and sequencing the fallen ssNA segment,
thereby allowing reconstruction of the track of the recoiling Au nucleus with nanometer accuracy.}
\label{fig6}
\end{figure}

\medskip\noindent
{\bf ``NA Tracking Chamber":}

The detector is modular and consists of a series of identical units stacked on top of each other.  It is like a book and the WIMP travels sequentially through the pages.  
Each module consists of the following layers.   On the top is a 1 $\mu$m layer of mylar (which is inactive from the point of view of incoming WIMPs).
Next is a 5-10 nanometer thick layer of gold, corresponding to roughly 10-25 atoms of Au in thickness.  It is with these Au nuclei that the WIMPs will mostly interact, since the atomic number for gold, and hence the WIMP spin-dependent interaction cross-section, is larger than for the organic atoms found in mylar: carbon, oxygen and hydrogen. The interaction of the WIMP with the Au nucleus will give it 
a kick of $\sim$ 10 keV out of the film.
 Hanging from the gold film is an ordered periodic array of ssNA strands, which can be thought of as a curtain of NA through which the recoiling Au nuclei will travel.  
 As the Au nucleus moves through the ``NA Tracking Chamber", it will break ssNA strands along its trajectory. More accurate studies of this breaking
 will be required, {\it e.g.} by calibrating the response of the ssNA to heavy ions of a given energy (such as  5, 10, 30 keV Ga ions that may be obtained from an ion implementation machine). 
 The required amount of energy to break the strand \footnote{Whereas a recoiling
 Au nucleus will easily break ssDNA, it would only nick dsDNA.  Thus in the current implementation we propose using single stranded nucleic acids (ssNA). However in future designs
 it may be useful to use a combination of the two.} is estimated to be 10 eV, but more accurate values must be
 obtained experimentally.  Thus we estimate that it will take hundreds to thousands of direct hits
 of Au on ssNA, corresponding to a comparable number of breaks of ssNA strands, to stop a gold nucleus.
 Currently off the shelf technology consists of arrays containing ssNA strands that are 250 bases in length (manufactured by Illumina Inc.).
 The average length of single-stranded NA is up to about 0.7 nm per base when fully stretched.
 Thus this corresponds to $\sim 100-200$ nm length NA strands.  Somewhat longer strands would be ideal as then
all the Au nuclei would be stopped in the NA Tracking Chamber and one could obtain the maximum information in reconstructing the particle's track. 
 
  The goal is to have the ssNA strands periodically ordered, with $\sim$ 10 nanometer distances between them.
  The NA can be immobilized at one end by a variety of means. For example, a Au-sulfur bond with NA terminally labeled with a thiol group \cite{pap5}. 
   Alternatively, Au coated with Streptavidin (a biotin-binding protein), will hold NA labeled with biotin.  Even simple positively-charged dots can be effective \cite{dbc}.  
   The challenge is to get single molecules attached to the gold film on a well defined two-dimensional grid, ``polka dot", pattern. Grid dots 5 nm in diameter can be 
   microfabricated with a spacing of 10 nm in the x- and y-directions, but to guarantee close to a single NA molecule per dot requires a trick like steric 
   hindrance (aka ``Polony exclusion principle") \cite{dbc} or designed 3D NA-nanostructures \cite{pap5}.  These two methods  can also help simplify manufacturing by 
   dividing the gold film into a 10nm grid from a 100 nm grid (the latter made by conventional photolithography or interference methods). If very non-repetitive
    NA curtains of known positioning and sequence are sought, then these can be constructed from synthetic NA selectively amplified from oligonucleotide 
    NA chips \cite{d} and/or certain natural genomes.
   
 The ssNA Tracker will operate with helium or nitrogen gas in between the hanging ssNA strands.   Oxygen in air would react and water would absorb too much energy.

Individual strands differ only in the ``terminus pattern" of say 20-100 bases at the bottom that identify the individual strands (more accurately members of a small bunch of NA strands).
   One may think of the Au film as a grid of squares that are 1$\mu$m $\times$ 1$\mu$m in size.  We will call the ssNA hanging down from one grid square a ``bundle" of NA.
All the ssNA strands have the same base sequence ordering, except the last 20-100
at the bottom are different for each grid area --- the same for all the ssNA within one bundle.  Thus one can localize the hit of the recoiling Au nucleus on the ssNA to 
1$\mu$m $\times$ 1$\mu$m in x-y.  
One can think of this as bar coding the ssNA strands (i.e. attaching tags of different colors to the bottoms of a group of NA strands) that cover a square region that is 1$\mu$ by 1$\mu$; thus the x-y resolution 
of the track will be micron-sized. 

The individual strands will be terminated with magnetic needles. 
Once a ssNA strand is severed, the segment falls to a collecting plate at the bottom.  A magnetizable rod that is 3-4 nm in diameter and 50 nm long
is attached to the bottom of the strand, provides the weight to pull it down, and is used to "scoop" the cut ssNA.
Roughly once an hour the ssNA segments are scooped up.  At that point they are amplified a billion fold using PCR, and then they are sequenced.  The location where
the NA was severed is identified, with nm resolution in z and micron resolution in x and y.  In this way the track of the recoiling Au nucleus from the WIMP interaction can
be reconstructed.

{\it Mass and Volume of Detector:} 
To study light WIMPs with $m_\chi \sim 10$ GeV, a single kilogram of gold is enough as target material at an energy threshold of 0.5 keV.   The most sensitive competing experiments, LUX and XENON, have very poor sensitivity to these low mass WIMPs, so that the current bounds on the WIMP scattering cross sections are not very strong.  Hence 1 kg of gold can be used to search for light WIMPs below the current bounds. Specifically, we find that we can use 5000 plates of gold, each $1$ m$^2$ in area,
separated by a micron layer of mylar, and containing a micron length of ssNA, totaling to a complete detector volume of $0.01$ m$^3$ in size.  

Studying WIMPs with $m_\chi \sim 100$ GeV will be much more difficult with our setup.  This is the mass where LUX and XENON are optimized, and presumably 
in their next runs the bounds will approach the neutrino background that also produces elastic scattering events and will make WIMP searches to lower cross sections very difficult.  We reiterate, however, that the ssNA tracker will do much better for light WIMPs than any of the xenon-based detectors, even without the search for directionality.  Xenon-based detectors can never probe WIMPs below 10 GeV: given the energy thresholds of the experiments, they would only be 
sensitive to particles moving at speeds faster than escape velocity from the Galaxy.  Additionally, the ssNA tracker again becomes competitive for high mass WIMPs 
$m_\chi > 500$ GeV (where the other experiments lose sensitivity).  

Importantly, there is no reason to exclusively use gold as the target element.  We have used gold for simplicity in this paper, since gold with NA attached
already exists and can be purchased off the shelf.  Instead we can use many other elements as foils in our detector.  Since attaching gold to ssNA
is well-studied, we can even use a single atom layer of gold (as the element that attaches to the ssNA) and then have a larger amount of
a different target material on top of the gold.  Again, these possibilities are probed in the Appendix.

{\it Diurnal Modulation:}
The ssNA tracker can be used to study diurnal modulation.  From the point of view of an experiment located at roughly 45 degrees latitude (such as from the vantage point of the SNOLAB in Sudbury, Canada, or the Sanford mine in South Dakota), at one point of the day, the WIMP wind is coming more or less from directly overhead.  Thus typical WIMPs will send the Au atoms in a direction parallel to the NA strands and will not break the NA as they traverse the detector (assuming the detector is parallel to the Earth).  However, as the Earth rotates, twelve hours later, the detector is now pointing 90 degrees away from the WIMP wind (see Figure 1). In this case there will be significant breakage of NA. As we will discuss in the following section, there is roughly 20 degree uncertainty in the direction of a typical Au nucleus ejected from the Au film; yet the differentiation in direction is much larger, roughly 90 degrees. Thus the day/night effect can be determined by the ssNA tracker, as the count rate goes through a daily maximum and minimum.

\medskip\noindent
{\bf Dynamics of the gold nucleus}

 SRIM  (the Stopping and Range of Ions in Matter) \cite{SRIM} is a simulation program that studies the interaction of a heavy ion moving through a specific medium.   
 We performed SRIM simulations to obtain quantitative estimates of the motion of the recoiling Au nucleus (from a WIMP interaction) as it moves through the detector. We performed this calculation for Au, but similar results are for any high Z and high density metals. Two stages were studied. First we
 studied  the interaction of the Au nucleus with other Au nuclei in the gold film, and obtained its average properties as it exits the gold film and enters the hanging curtain 
 of ssNA strands.  Second, we studied the interaction of the Au nuclei with the strands of ssNA it encounters, and estimated the distance the Au travels before it 
 comes to a stop.  We performed these studies for two different WIMP masses, 200 GeV and 1 TeV.  SRIM incorporates the Lindhard equation,
 Bethe-Block equation, and experimental results on those ion/medium interactions that have been measured. The initial estimates we obtaining using SRIM
 of course must be tested with real data in the future. 

  In using SRIM we considered a 5nm thick sheet of gold in a detector module, and we investigated
  the result of WIMP/nucleus interaction at the midpoint of 2.5 nm from the surface of the sheet.   
 To be quantitive, we assumed a typical energy recoil $E_{nr}=\frac{1}{2}E_{max}=\frac{\mu^2}{M}v^2$ in the interaction, 
 where we have used Eq[\ref{Emax}] in the previous expression. For the WIMP velocity we used the mean speed of a Standard Halo Model distribution in the frame of the laboratory, i.e. $v=300$ km/sec.   We took the direction of the recoiling Au nucleus to be perpendicular to the Au foil. 
  For a 200 GeV WIMP, we found the following results.  The recoil energy of a typical Au nucleus due to a WIMP interaction
 is 50 keV.  The Au nucleus then encounters other Au nuclei on its way out of the Au foil;  but the mean stopping range would be 7.3 nm, so that the typical Au 
 nucleus does escape the foil.  At this point its energy is roughly halved, at 27.5 keV, and its angle is 17.1$^\circ$ off from its incident angle (perpendicular
 to the foil).  Before stopping,
 the Au nucleus traverses approximately 2.55 $\mu$m of ssNA, which translates to about 255 strands crossed.
  Alternatively, for a 1 TeV WIMP,  the recoil energy given to the Au nucleus by the WIMP interaction is 131 keV, the mean stopping range in gold would be 14 nm,
 the Au nucleus typically leaves the foil with 110 keV and an angle of 10.9$^\circ$, and it then stops after traversing 5.83 $\mu$m of ssNA, which translates to about 583 strands crossed.  Our calculation is based on the assumption that the ssNA strands are simple packed next to each other with a separation of 10 nm and the ''NA tracking chamber" is optically thick. We  conclude that the number of ssNA strands crossed by a gold nucleus is on the order of hundreds of ssNA strands. 
 
 The original recoiling Au nucleus, which scattered off the galactic WIMP will also interact with other nuclei in the gold film. As the original Au nucleus moves through the film, it will cause a fraction of other gold nuclei to escape into the ssNA tracking chamber. A gold nucleus interacting with a 200 GeV (1000 GeV) WIMP will induce an average of 21.3 (21.4) Au atoms with 63 eV (89 eV) each to traverse around 18 (20) ssNA strands. The number of strands crossed by the secondary ejected gold atoms is much less compared to the hundreds of strands crossed by the scattered gold nucleus from the WIMP collision.

\medskip\noindent
{\bf First Generation Implementation:}

The initial goal will be to identify a head/tail asymmetry of the WIMPs. As described above, the number of WIMPs coming from the direction 
of Cygnus should be ten times that from the opposite direction, since we are moving into the Galactic wind of WIMPs. Merely identifying this head/tail
asymmetry may be enough to argue (together with annual modulation) that WIMPs have been discovered.   
The design described above automatically provides this distinction.  WIMPs that come in the direction of first passing through the mylar, then into the Au film, 
and then into the ssNA strands will be detected.  However, WIMPS that go the other way will not:  these first go unnoticed through the NA, then interact in the gold,
producing recoiling gold nuclei that simply stop in the mylar. This is a simple implementation of head/tail differentiation.  Then the entire detector may be flipped 180 degrees, so
that it is sensitive to only WIMPs coming from the other direction.  Note that in this case, we do not need the ordered array of ssNA- only the number of broken ssNA is important. 
Once this simplest version works, the goal of the next generation detectors will be to look for the actual track of the recoiling nucleus with nanometer resolution, 
using longer ssNA strands in a periodic array.

\medskip\noindent
{\bf Background rejection in the proposed Au/ssNA detector:}  

There are many sources of background that could mimic a WIMP signal.  The improved granularity of our detector --- nanoscale vs. micron length scale --- should
help with background rejection.  In previous detector development, the background rejection has scaled with the spatial resolution, so one might
hope that the background rejection improves a billion fold in these new detectors; but this must be verified.

  Naturally occurring DNA (and RNA)  itself contains radioactive C$^{14}$ and K$^{41}$.  
 The DNA (RNA) in the detector must be made of old carbon, and potassium can be replaced 
 by other moieties, as discussed further at the end of this section.   Studies must be performed of the
 radioactivity of thin films of Au or other elements.

   Backgrounds that could be confused with WIMPs include $\gamma$, $\alpha$, electrons, and cosmic rays (CR).  
  Yet the ranges of these particles in our detector will be at least 100 times as long as the range of a recoiling nucleus from a WIMP, so that the differentiation between them
  should easily be possible. The background with the shortest range is the $\alpha$ particle due to its relative larger mass and charge. A typical radioactive decay produces a 5 MeV $\alpha$ particle. The stopping distance of an $\alpha$ particle can be approximated with the ASTAR program. Using the ASTAR program, a 5 MeV $\alpha$ particle will traverse approximately 30 $\mu$m (10 $\mu$m) in mylar (gold) \cite{ASTAR}. In contrast, SRIM simulations estimate that the stopping distance of a 10 keV gold nucleus in mylar is 20 nm \cite{SRIM}.       
  Whereas recoiling Au nuclei will be stopped in a single module of Au foil + ssNA, the background particles will travel much farther as they traverse and interact with many
  of the sequential modules.  We have described our detector design as a book with many pages.  The recoiling nuclei from WIMP interactions will never make
  it past the first page, whereas the $\gamma$, $\alpha$, electrons, and cosmic rays (CR) will go through multiple pages.
Thus the signature of a WIMP is that there is an interaction in ``one and only one" gold film.  

In addition, as discussed, background signals are expected to be isotropic whereas the
WIMPs are expected to exhibit head/tail as well as temporal asymmetries.
  
  Even more accurate determinations can be made by measuring the values of dE/dx of the various particles traveling through the detector, as 
  this will allow differentiating between them.
  Such a measurement requires the spatial resolution to be shorter than the range of the particle.
  Previous to the use
  of NA, the spatial resolution of detectors wasn't good enough to make this determination, so that the best that could be done was the measurement of the 
integrated value of dE/dx over the spatial resolution of the detector.  The NA tracking chamber provides for the first time the capability of obtaining this quantity.
  
  As in all WIMP direct detection experiments, once these backgrounds are removed, the chief remaining troublesome particles that can mimic a WIMP signal are fast neutrons.  
    Further studies, both experimental tests and Monte Carlos, will
be required to better understand the neutron background.
The capability of tracking particles as they move through the ssNA should assist with this distinction.

{\it Cost Estimates using RNA:}
The detector could be designed using either DNA or RNA.  Since the cost per kg of the least expensive nucleic acid is yeast RNA (\$750/kg),
we will here give cost estimates using RNA rather than DNA.  We can make $^{40}$K and $^{14}$C- depleted bacterial 23S-rRNA (2904 nucleotides long) for about \$8000/kg. Rather than using weights at the bottom (which does not allow dense packing), we will keep the nucleic acid straight (rather than curling up) by using short oligonucleotides (10-mers) complementary to the 5' and 3' ends of the RNAs and immobilized on the surfaces of 25*25*0.1 mm coverslips. A third 10-mer (fluorescently labeled) will be located 200 nm from the end of the 1400 nm long stretched rRNA. The 200 nm is chosen to be out of range of TIRF, so when an RNA is cleaved the fluor moves to $<$1nm and can be detected using TIRF. The three 10-mers will constitute 1\% of the mass. To get to 1 kg would require 10 million coverslips scanned at 120 per second to finish one scan cycle per day.

 Here we estimate the number of radioactive decay events in 0.1 kg of RNA and show that identifying them is financially affordable; the decay numbers
 and cost scale linearly with the amount of nucleic acid.
 We can make the RNA $^{40}$K depleted by using Na$^+$ as the counter-ion for the RNA instead of K$^+$ during
purification. We can make the RNA $^{14}$C-depleted by using petrochemicals (e.g. methane) as the sole carbon
source for bacterial growth. We may assume a ratio of 10 million for Na:K and a $^{40}$K  natural abundance of 0.012\%
and half-life of 1.25 $\times 10^9$ years. For 0.1 kg we have 2 $\times 10^{23}$ molecules of nucleotides = 2 $\times 10^{11}$atoms of $^{40}$K. Thus, we
expect 6 decays per 0.1 kg of RNA per day. Each nucleotide molecule contains 9.5 carbon atoms on average and the $^{14}$C
half life is 5,730 years.
In the Borexino Counting Test Facility, a $^{14}$C/$^{12}$C ratio of 1.94 $\times 10^{-18}$ has been determined for 
organic syntillation fluid composed of old carbon derived from petroleum \cite{DNAimpure}. Extrapolating to  a detector composed of 0.1 kg of nucleic acid, 
this low concentration of C$^{14}$  yields $\sim 2$ decay per day, which can be subsequently rejected due to the multi-module (Au foil + ssNA) interaction of the background.  
Fortunately, biological technology has advanced to such an extent that it is currently possible to sequence millions of nucleic acid strands a day for  $\sim$ \$1,000.
Background events are analyzed cheaply due to the affordable price of sequencing. 

\section{\label{sec:Summary} Summary}

A major step forward in the field of direct detection would be the development of 
detectors with directional capability.  By contrasting the count rates in a detector in the direction toward and away from the Galactic WIMP ``wind" that the 
Sun is moving into,
the statistical requirements on the number of detected WIMPs drops to $\sim 100$ rather than thousands without the directional sensitivity.
In the paper we proposed using DNA or RNA as a detector material that can provide nanometer resolution tracking.  
We presented a particular design consisting of modules of thin gold films with single stranded NA hanging down from each film.
The NA strands all consist of (almost) identical sequences of bases (combinations of A,C,G,T), with an order that is well known.
An incoming WIMP from the Halo of our Galaxy strikes one of the gold nuclei and knocks it out of the film with $\sim$ 10 keV
of energy.  The Au nucleus moves forward into the strands of NA, traverses hundreds of these strands, and whenever it hits one, breaks the ssNA.  
The locations of the breaks are easy to identify, using PCR to amplify the broken segments a billion fold followed by NA sequencing to locate the break.
In this way the path of the recoiling nucleus can be tracked to nanometer accuracy.  

We note that this design is not restricted to the use of a particular element, e.g. Au nuclei, which can be interchanged with many different nuclei;
for example one may use those with high atomic number so as to maximize the
SI interaction rate, or those with low atomic number to study the lightest WIMPs.  By using a variety of different materials, it should be possible to identify the mass and cross-section of the interacting WIMP.  In addition,
although the specific detector design may be modified, the important new development 
is the idea of using NA in lieu of other detector materials to provide better tracking resolution so that directionality of the WIMPs can be determined.  More generally, it is easy to imagine multiple applications for nanometer tracking beyond that of WIMP detection.


\begin{acknowledgments}
  K.F.\  and A.L.\ acknowledge the support of the DOE and the Michigan
  Center for Theoretical Physics via the University of Michigan.
  K.F.\ thanks the Caltech Physics Dept for hospitality during her visit.
  We are grateful to Dave Gerdes, Rachel Goldman, Sharon Glotzer, Cagliyan Kurdak, Chris Meiners, Joanna Millinchuk, Neelima Sehgal,
  and Greg Tarle for useful conversations.  AKD would like to thanks L. Stodolsky, F.T. Avignone, S. Nussinov, G. Gelmini, R. Bernabei, J.P. Belli, B. Cabrera, B. Sadoulet  and Munich group for years of discussions about �what to do about DM problem�.   
 
\end{acknowledgments}




\appendix*
\section{}
\indent In this paper we have argued that a variety of elements could be used as the detector target material. In particular, choosing an element with a similar mass as the WIMP will be kinematically favored. The recoil energy after a WIMP/nucleus elastic collision is proportional to $E_{nr}\propto \mu^2/M$; where $\mu=\frac{m_\chi M}{m_\chi+M}$, $m_\chi$ is the mass of the WIMP and $M$ is the mass of the target element. Thus, the recoil energy imparted on the recoiled nucleus is maximized when $M=m_\chi$. In this appendix we study a number of potential elements for a specific WIMP mass range. Specifically, we choose the following materials: 1) tungsten (W) for a WIMP mass $m_\chi\cong 200$ and $400$ GeV, 2) tin (Sn) for $m_\chi \cong 100$ GeV, and 3) copper (Cu) for $m_\chi \cong 50$ GeV. All of these proposed target elements are fairly inexpensive.
\\
\indent In the table below, we make several calculations in order to establish the validity of each assumed element as a target material for their respective WIMP mass range. First, we calculate the typical recoil energy of the scattered nucleus and its stopping distance as it moves through an infinite slab of the same element ($r_{stop}$). The typical recoil energy is given by $E_{nr}=\frac{1}{2}E_{max}=\frac{\mu^2}{M}v^2$, where the WIMP velocity is taken to be $v=300$ km/s. Given the recoiled energy, the stopping distance is calculated utilizing the SRIM program. Secondly, we establish the minimum velocity ($v_{min}$) of the WIMP that would produce a scatter in our detector. As well, we utilize Eq [\ref{eqn:dRdEnr}] in order to determine the expected number of WIMP/nucleus scattering events per kg per year ($N_{e}$), and the amount of target mass ($M_{T}$) needed to have 20 (50) events in a 3 year run assuming the LUX spin-independent cross-section limit. In order to make the previous calculations ($v_{min}$, $N_{e}$ and $M_{T}$), we need to assume a realistic energy threshold. Since the design, geometry, and choice of material influence the energy threshold ($E_{th}$) of the proposed detector, then calculations will be made assuming: 1) $E_{th}=0.5$ keV, 2) $E_{th}=5$ keV and 3) $E_{th}=10$ keV. As well, form factors have been ignored; thus our results for $N_e$ and $M_T$ are only approximate. The table below shows the results obtained. 
\\

\providecommand{\tabularnewline}{\\}

\makeatother

\begin{tabular}{|l|l|l|l|l|l|}

\cline{2-6} 
\multicolumn{1}{l|}{} & \multicolumn{1}{l|}{$m_{\chi}$ {[}GeV{]}} & 400 & 200 & 100 & 50\tabularnewline
\cline{2-6} 
\multicolumn{1}{l|}{} & Target Element & W  & W & Sn  & Cu \tabularnewline
\cline{2-6} 
\multicolumn{1}{l|}{} & Mass of target {[}GeV{]} & 171.3 & 171.3 & 110.6 & 59.2\tabularnewline
\cline{2-6} 
\multicolumn{1}{l|}{} & $E_{nr}$ {[}keV{]} & 84.0 & 49.7 & 24.9 & 12.4\tabularnewline
\cline{2-6} 
\multicolumn{1}{l|}{} & $r_{stop}$ {[}nm{]} & 10.2 & 7.1 & 11.4 & 5.6\tabularnewline
\hline 
 & $E_{th}=10$ keV & 16.4 & 21.3 & 30.0 & 42.6\tabularnewline
\cline{2-6} 
$v_{min}$ {[}km/s{]} & $E_{th}=5$ keV & 51.8 & 67.3 & 95.0 & 134.6\tabularnewline
\cline{2-6} 
 & $E_{th}=10$ keV & 73.2 & 95.2 & 134.3 & 190.4\tabularnewline
\hline 
 & $E_{th}=0.5$ keV & 20.7 & 7.4 & 0.8 & 0.17\tabularnewline
\cline{2-6} 
$N_{e}$ {[}events/kg/year{]} & $E_{th}=5$ keV & 20.2 & 7.1 & 0.74 & 0.15\tabularnewline
\cline{2-6} 
 & $E_{th}=10$ keV & 19.7 & 6.8 & 0.68 & 0.12\tabularnewline
\hline 
 & $E_{th}=0.5$ keV & 0.32 & 0.90 & 8.3 & 39.2\tabularnewline
\cline{2-6} 
$M_{T}$ for 20 counts {[}kg{]} & $E_{th}=5$ keV & 0.33 & 0.94 & 9.0 & 44.4\tabularnewline
\cline{2-6} 
 & $E_{th}=10$ keV & 0.34 & 0.98 & 9.8 & 55.6\tabularnewline
\hline 
 & $E_{th}=0.5$ keV & 0.81 & 2.3 & 20.8 & 98.0\tabularnewline
\cline{2-6} 
$M_{T}$ for 50 counts {[}kg{]} & $E_{th}=5$ keV & 0.83 & 2.4 & 22.5 & 111\tabularnewline
\cline{2-6} 
 & $E_{th}=10$ keV & 0.85 & 2.5 & 24.5 & 139\tabularnewline
\hline 

\end{tabular}
\\
\\
TABLE 1: This table shows the result of using different materials as target elements depending on the WIMP mass range of interest. We make the following choices of target material: tungsten for WIMP mass $m_\chi=400$ and 200 GeV, tin for $m_\chi=100$ GeV, and copper for a $m_\chi=5$ GeV WIMP. In particular, Table 1 shows the average recoil energy ($E_{nr}$) and stopping distance ($r_{stop}$) for a recoiled nucleus (tungsten, tin and copper) after elastically scattering with a massive WIMP (400 and 200 GeV, 100 GeV and 50 GeV respectively). We also show the minimum velocity ($v_{min}$) the galactic WIMP needs in order to produce a signal above the assumed realistic energy threshold ($E_{th}= 0.5$ keV, $5$ keV and $10$ keV). Next, we calculate the number of events per kilogram per year ($N_e$) expected for each element and corresponding WIMP interacting with a spin-independent cross-section amplitude set at the LUX upper bound. The last two rows give the total amount of target material $M_T$ necessary to get 20 and 50 WIMP/nucleus events, respectively, after operating the detector for 3 years. 
\\
\\
\\
\indent We note that it is possible to look for high mass WIMPs ($m_\chi \geq 200$ GeV) with about 1 kg of tungsten as the target element of the detector. One kilogram of tungsten is roughly enough to get 50 events in three years of operationally running the detector at an energy threshold $E_{th} < 10$ keV. In contrast, WIMPs of mass close to 50 GeV are harder to detect and suggest having a detector of around 100 kg or more target mass. The ability to switch between different target elements for the proposed detector gives it versatility to probe different WIMP mass ranges and study different aspects of the WIMP/nucleus cross-section (e.g. proton/neutron-WIMP cross-section). 

\end{document}